\documentclass[aps,pra,superscriptaddress,floatfix,showpacs,10pt,twocolumn]{revtex4}
\usepackage[dvips]{graphicx}
\usepackage{amsmath}

\begin{document}

\title{Quantum Information Processing using coherent states in cavity QED}
\author{Ming Yang}
\email{mingyang@ahu.edu.cn}
\author{Zhuo-Liang Cao}
\email{zlcao@ahu.edu.cn} \affiliation{School of Physics {\&}
Material Science, Anhui University, Hefei, 230039, PRChina}

\begin{abstract}
Using the highly detuned interaction between three-level
$\Lambda$-type atoms and coherent optical fields, we can realize
the C-NOT gates from atoms to atoms, optical fields to optical
fields, atoms to optical fields and optical fields to atoms. Based
on the realization of the C-NOT gates we propose an entanglement
purification scheme to purify a mixed entangled states of coherent
optical fields. The simplicity of the current scheme makes it
possible that it will be implemented in experiment in the near
future.
\end{abstract}
\maketitle

Entanglement plays a key role not only in refuting the local
''hidden variable'' theory~\cite{Einstein:1935, Bell:1965} but in
quantum information processing also, such as quantum
teleportation~\cite{Bennett:1993, Bouwmeester:1997}, quantum dense
coding~\cite{Bennett:1992, Mattle:1996}, quantum
cryptography~\cite{Ekert:1991, Bennett:1994} and so on.

The preparation of entangled states becomes a vital step in
quantum information processing (QIP). Recently, the generation of
entanglement has been realized by NMR~\cite{generation:nmr1,
generation:nmr2}, SPDC~\cite{Dik:1999, generation:spdc}, Cavity
QED~\cite{generation:qed1, generation:qed2}, and Ion
Trap~\cite{generation:ion1} schemes. The generation scheme for
polarization entangled photon state using SPDC has been
reported~\cite{Dik:1999}, and it has been realized in experiment.
Now, the techniques for generating entangled photon pairs have
become rather mature. At the same time, quantum teleportation of
unknown photon state and quantum cryptography process are all
realized in experiment~\cite{Bouwmeester:1997}. But, for the
entangled atomic state, it is not the case. Although many
theoretical schemes for the generation of entangled atomic state
have been proposed, the number of the schemes that can really be
realized in experiment is very small. Hitherto, only the
entanglement of two atoms has been realized
experimentally~\cite{generation:qed2, generation:qed3}. The
teleportation and cryptography schemes can not yet been
implemented in experiment for atoms. To realize the atom-based
quantum information processing, we must present more
experimentally efficient atom-base QIP schemes.

In view of the previous QIP schemes, we found that the quantum
controlled-not (C-NOT) gate is the key part of a total scheme.
C-NOT gate can not only generate entangled states but realize the
teleportation process through Bell state measurement also. In the
cavity QED domain, the schemes for C-NOT gate have been proposed.
By far the most efficient scheme is the one proposed by
Zheng~\cite{generation:qed2}, where the interaction between two
atoms induced by a dispersive cavity mode plays a key role and the
C-NOT operation from atom to atom has been realized.

In this contribution, we will propose an efficient scheme to
realize the C-NOT gates between atom and field of coherent light.
The C-NOT operations involve four kinds: C-NOT gate from atom to
field of coherent light, C-NOT gate from coherent light to atom,
C-NOT gate from one atom to another and C-NOT gate from one field
to another. Being more efficient than the proposal of
Zheng~\cite{generation:qed2}, our scheme not only can generate
multi-atom entangled states but also can generate multi-mode
entangled coherent states. The scheme is mainly based on the
dispersive interaction between atoms and coherent optical fields.

Consider the interaction between an $\Lambda$-type three-level
atom and a coherent optical field. Here the two lower levels of
the atom are degenerate, and the frequency of the coherent optical
field $\omega_{c}$ is largely detuned from the atomic transition
frequency $\omega_{0}$ between the degenerate lower levels and the
upper level. In this large detuning limit, the upper level
$|i\rangle$ can be adiabatically eliminated during the
interaction.

Then the effective Hamiltonian for the system can be expressed as
follow~\cite{xu}:

\begin{equation}
\hat{H}=-\lambda a^{+}a(|e\rangle \langle g|+|g\rangle \langle
e|)-a^{+}a(\beta _{1}|e\rangle \langle e|+\beta _{2}|g\rangle
\langle g|)  \label{hamiltonian}
\end{equation}
where$\lambda =g_{1}g_{2}/\Delta, \beta_{1}=g_{1}^{2}/\Delta,
\beta_{2}=g_{2}^{2}/\Delta,$ with $\Delta=\omega_{0}-\omega_{c} $
being the detuning between atomic transition frequency and the
frequency of the coherent light, $g_{1},g_{2}$ being the coupling
constant between the cavity mode and the transitions $|i\rangle
\rightarrow |e\rangle, |i\rangle \rightarrow |g\rangle$
respectively. Suppose $g=g_{1}=g_{2}$, $\lambda
=\beta_{1}=\beta_{2}=g^{2}/\Delta$.

Suppose that the atom is initially prepared in $|e\rangle $ state,
and the optical field is in coherent state. Then the interaction
between atom and coherent optical field will lead to the following
evolution:

\begin{equation}
|e\rangle |\alpha \rangle \overset{U(t)}{\longrightarrow
}(1/2)[(|\alpha \rangle +|\alpha e^{2i\lambda t}\rangle)|e\rangle
-(|\alpha \rangle -|\alpha e^{2i\lambda t}\rangle)|g\rangle]
\label{evolution1}
\end{equation}
Similarly, if the atom is initially prepared in $|g\rangle $ state, the
evolution takes a new form:

\begin{equation}
|g\rangle |\alpha \rangle \overset{U(t)}{\longrightarrow
}(1/2)[(|\alpha \rangle +|\alpha e^{2i\lambda t}\rangle)|g\rangle
-(|\alpha \rangle -|\alpha e^{2i\lambda t}\rangle)|e\rangle]
\label{evolution2}
\end{equation}
If we select the atomic velocity to realize the interaction time
$t=\pi/(2\lambda)$, then Eqs (\ref{evolution1}, \ref{evolution2})\
will become:
\begin{subequations}
\begin{equation}
|e\rangle |\alpha \rangle \overset{U(t)}{\longrightarrow
}(1/2)[(|\alpha \rangle +|-\alpha \rangle)|e\rangle-(|\alpha
\rangle -|-\alpha \rangle)|g\rangle]
\end{equation}
\begin{equation}
|g\rangle |\alpha \rangle \overset{U(t)}{\longrightarrow
}(1/2)[(|\alpha \rangle +|-\alpha \rangle)|g\rangle -(|\alpha
\rangle -|-\alpha \rangle)|e\rangle]
\end{equation}
\end{subequations}
Let$|\alpha_{+}\rangle =(1/\sqrt{2})(|\alpha \rangle +|-\alpha
\rangle)$, $|\alpha_{-}\rangle =(1/\sqrt{2})(|\alpha \rangle
-|-\alpha \rangle) $, then we get
\begin{subequations}
\begin{equation}
|e\rangle |\alpha \rangle \overset{\lambda
t=\pi/2}{\longrightarrow }(1/\sqrt{2})[|\alpha _{+}\rangle
|e\rangle -|\alpha _{-}\rangle |g\rangle] \label{detailed
evolution1a}
\end{equation}
\begin{equation}
|g\rangle |\alpha \rangle \overset{\lambda
t=\pi/2}{\longrightarrow }(1/\sqrt{2})[|\alpha _{+}\rangle
|g\rangle -|\alpha _{-}\rangle |e\rangle]  \label{detailed
evolution1b}
\end{equation}
From the analysis of coherent state~\cite{van}, we get that :
$|u\rangle =1/\sqrt{2(1+e^{-2\|\alpha|^{2}})}(|\alpha \rangle
+|-\alpha \rangle)$, $|v\rangle =1/\sqrt{2(
1-e^{-2|\alpha|^{2}})}(|\alpha \rangle -|-\alpha \rangle)$ are two
orthogonal basis. Here we use $|\alpha_{+}\rangle $ and
$|\alpha_{-}\rangle $ to replace $|u\rangle $ and $|v\rangle $. In
fact, when $|\alpha|=3$, the approximation is rather perfect. So
the entangled states in Eqs (\ref{detailed evolution1a},
\ref{detailed evolution1b}) are maximally entangled states between
the atom and the coherent optical field.

Next, we will give the evolution of the system for different initial state:
\end{subequations}
\begin{subequations}
\begin{equation}
|e\rangle |-\alpha \rangle \overset{\lambda
t=\pi/2}{\longrightarrow}(1/\sqrt{2})[|\alpha_{+}\rangle |e\rangle
+|\alpha_{-}\rangle |g\rangle] \label{detailed evolution2a}
\end{equation}
\begin{equation}
|g\rangle |-\alpha \rangle \overset{\lambda
t=\pi/2}{\longrightarrow }(1/\sqrt{2})[|\alpha_{+}\rangle
|g\rangle +|\alpha_{-}\rangle |e\rangle] \label{detailed
evolution2b}
\end{equation}
\end{subequations}
From Eqs (\ref{detailed evolution1a}, \ref{detailed evolution1b},
\ref{detailed evolution2a}, \ref{detailed evolution2b}), we can
give the following operations:
\begin{subequations}
\begin{equation}
|\alpha_{+}\rangle |e\rangle \longrightarrow |\alpha_{+}\rangle
|e\rangle \label{cnot1a}
\end{equation}
\begin{equation}
|\alpha_{+}\rangle |g\rangle \longrightarrow |\alpha_{+}\rangle
|g\rangle \label{cnot1b}
\end{equation}
\begin{equation}
|\alpha_{-}\rangle |e\rangle \longrightarrow -|\alpha_{-}\rangle
|g\rangle \label{cnot1c}
\end{equation}
\begin{equation}
|\alpha_{-}\rangle |g\rangle \longrightarrow -|\alpha_{-}\rangle
|e\rangle \label{cnot1d}
\end{equation}
\end{subequations}
which are C-NOT operations from optical field to atom, and the
operations:
\begin{subequations}
\begin{equation}
|-\rangle |\alpha \rangle \longrightarrow |-\rangle |\alpha \rangle
\label{cnot2a}
\end{equation}
\begin{equation}
|-\rangle |-\alpha \rangle \longrightarrow |-\rangle |-\alpha \rangle
\label{cnot2b}
\end{equation}
\begin{equation}
|+\rangle |\alpha \rangle \longrightarrow |+\rangle |-\alpha \rangle
\label{cnot2c}
\end{equation}
\begin{equation}
|+\rangle |-\alpha \rangle \longrightarrow |+\rangle |\alpha \rangle
\label{cnot2d}
\end{equation}
\end{subequations}
which are C-NOT operations from atom to optical field. Where
$|+\rangle =(1/\sqrt{2})(|e\rangle +|g\rangle)$, $|-\rangle
=(1/\sqrt{2})(|e\rangle -|g\rangle)$.

With the C-NOT gates being realized, we can realize the generation
of entangled atomic states and entangled coherent states. In
addition, we also can realize the purification of the mixed
entangled atomic states and mixed entangled coherent states.

Firstly, we will consider the generation of maximally entangled
atomic states. After the first atom $|e_{1}\rangle $ has
interacted with the coherent optical field for
$t_{1}=\pi/(2\lambda)$, the evolution of the system can be
described by Eq (\ref{detailed evolution1a}). Then the second
atom, initially prepared in $|e_{2}\rangle $ state, will be sent
through the field area. If the interaction time is still
$t_{2}=\pi/(2\lambda)$, the total evolution reads:

\begin{eqnarray}
&&|\alpha \rangle |e_{1}\rangle |e_{2}\rangle \overset{\lambda
t_{1}=\pi/2}{\longrightarrow}(1/\sqrt{2})[|\alpha_{+}\rangle
|e_{1}\rangle |e_{2}\rangle -|\alpha _{-}\rangle
|g_{1}\rangle |e_{2}\rangle]\nonumber\\
&&\overset{\lambda t_{2}=\pi/2}{\longrightarrow
}(1/\sqrt{2})[|\alpha_{+}\rangle |e_{1}\rangle |e_{2}\rangle
+|\alpha _{-}\rangle |g_{1}\rangle |g_{2}\rangle]
\label{geneatom1}
\end{eqnarray}
That is to say, after interactions the state of total system
becomes:
\begin{equation}
|\Psi _{total}\rangle \longrightarrow (1/\sqrt{2})[|\alpha \rangle
|\Phi _{12}^{+}\rangle +|-\alpha \rangle |\Phi _{12}^{-}\rangle]
\label{geneatom2}
\end{equation}
where $|\Phi_{12}^{+}\rangle =(1/\sqrt{2})[|e_{1}\rangle
|e_{2}\rangle +|g_{1}\rangle |g_{2}\rangle]$,
$|\Phi_{12}^{-}\rangle=(1/\sqrt{2})[|e_{1}\rangle
|e_{2}\rangle-|g_{1}\rangle |g_{2}\rangle]$ are two Bell states
for atoms $1$ and $2$.

Then we will detect the optical field. If the result is
$|\alpha\rangle$, the two atoms will be left in Bell state $|\Phi
_{12}^{+}\rangle $; If the result is $|-\alpha \rangle $, the two
atoms will be left in Bell state $|\Phi_{12}^{-}\rangle$.

In fact, if we do not detect the optical field after the second
atom flying out of the cavity, multi-atom entangled states can be
created. We will send the next atom $(|e_{n}\rangle)$ through the
field area after the previous one flying out of it. Then after the
last atom flying out of the cavity field, the optical field will
be detected. Conditioned on different results, the $n$ atoms will
be left in different maximally entangled states:
\begin{widetext}
\begin{eqnarray}
|\Psi _{n}\rangle
&&\longrightarrow(1/\sqrt{2})[(1/\sqrt{2})(|e_{1}\rangle
|e_{2}\rangle \cdots |e_{n}\rangle +(-1)^{n}|g_{1}\rangle
|g_{2}\rangle \cdots
|g_{n}\rangle)|\alpha \rangle \nonumber \\
&&+(1/\sqrt{2})(|e_{1}\rangle |e_{2}\rangle \cdots |e_{n}\rangle
+(-1)^{(n-1)}|g_{1}\rangle |g_{2}\rangle \cdots
|g_{n}\rangle)|-\alpha \rangle] \label{geneatomn}
\end{eqnarray}
\end{widetext}
Secondly, we will consider the generation of maximally entangled
coherent states. Let an atom, initially prepared in $|e\rangle$
state, interact with the first coherent optical field. Let the
interaction time satisfy $t_{1}=\pi/(2\lambda)$. After flying out
of the first cavity, the atom will be sent through the second
cavity field, and the interaction time is still
$t_{2}=\pi/(2\lambda)$. The evolution of the total system is:
\begin{widetext}
\begin{equation}
|e\rangle |\alpha _{1}\rangle |\alpha _{2}\rangle \overset{\lambda
t_{1}=\pi/2}{\longrightarrow }(1/\sqrt{2})[( |-\rangle |\alpha
_{1}\rangle +|+\rangle |-\alpha _{1}\rangle) |\alpha
_{2}\rangle]\overset{\lambda
t_{2}=\pi/2}{\longrightarrow}(1/\sqrt{2})[|-\rangle|\alpha_{1}\rangle
|\alpha _{2}\rangle +|+\rangle |-\alpha _{1}\rangle |-\alpha
_{2}\rangle]\label{genefield1}
\end{equation}
\end{widetext}
which can be expressed in another form:
\begin{widetext}
\begin{equation}
|\Psi _{total}\rangle \longrightarrow
(1/\sqrt{2})[(1/\sqrt{2})(|\alpha _{1}\rangle |\alpha _{2}\rangle
+|-\alpha _{1}\rangle |-\alpha _{2}\rangle) |e\rangle
+(1/\sqrt{2})(|\alpha _{1}\rangle |\alpha _{2}\rangle -|-\alpha
_{1}\rangle |-\alpha _{2}\rangle) |g\rangle] \label{genefield2}
\end{equation}
\end{widetext}
If we measure the state of the atom in basis ${|e\rangle
,|g\rangle}$, we can get the maximally entangled state of the two
cavity fields: $(1/\sqrt{2})(|\alpha_{1}\rangle |\alpha
_{2}\rangle+|-\alpha _{1}\rangle |-\alpha _{2}\rangle)$ for result
$|e\rangle$, $(1/\sqrt{2})(|\alpha _{1}\rangle |\alpha _{2}\rangle
-|-\alpha _{1}\rangle |-\alpha _{2}\rangle)$ for $|e\rangle$,

Like the generation of $n$-atom maximally entangled state, after
the atom flying out of the second cavity, we will not measure the
atomic state. Instead, we will send it through other coherent
optical fields one by one. In each cavity, the interaction time
are all equal to $t=\pi/(2\lambda)$. Then we can get the
$n$-cavity maximally entangled states:
\begin{widetext}
\begin{eqnarray}
|\Psi_{n}\rangle &\longrightarrow
&(1/\sqrt{2})[(1/\sqrt{2})(|\alpha_{1}\rangle |\alpha_{2}\rangle
\cdots |\alpha_{n}\rangle+|-\alpha_{1}\rangle
|-\alpha_{2}\rangle \cdots |-\alpha_{n}\rangle) |e\rangle \nonumber\\
&&+(1/\sqrt{2})(|\alpha_{1}\rangle|\alpha_{2}\rangle \cdots
|\alpha_{n}\rangle-|-\alpha_{1}\rangle|-\alpha_{2}\rangle \cdots
|-\alpha _{n}\rangle) |g\rangle]\label{genefieldn}
\end{eqnarray}
\end{widetext}
Due to cavity decay, the two-mode maximally entangled coherent
state in Eq (\ref{genefield2}) more easily evolve into a mixed
state. So next we will consider the purification of the mixed
entangled coherent state~\cite{bennett, pan, me}.

Suppose we have generated two pairs of the two-mode entangled
states of optical fields, and cavities $1$, $3$ are in the access
of one user Alice, cavities $2$, $4$ are in the access of the
other user Bob. At each side, there will be an auxiliary atom,
denoted by $a$ or $b$. With the help of the Ramsey Zones between
the two cavities. We can realize the C-NOT operations from cavity
$1$ to cavity $3$, and from cavity $2$ to cavity $4$.

Here atoms $a$, $b$ are all prepared at $|e\rangle $ state, and
the Ramsey Zones all have the same function, i.e. it can realize
the following rotations: $|e\rangle\rightarrow|+\rangle
=(1/\sqrt{2})(|e\rangle +|g\rangle)$,
$|g\rangle\rightarrow|-\rangle =(1/\sqrt{2})(|e\rangle -|g\rangle)
$. Then consider the purification procedure. The auxiliary atom
$a$ will be sent through the cavity $1$, Ramsey Zone $R_{a}$\ ,
and cavity $3$ one after another. The interaction times between
atom $a$ and cavity $1$, cavity $3$ are all equal to
$t=\pi/(2\lambda)$. Then we find that, if the state of cavity $1$
is expressed in $\{|\alpha_{+}\rangle ,|\alpha_{-}\rangle\}$
basis, and the state of cavity $3$ is expressed in $\{|\alpha
\rangle ,|-\alpha \rangle\}$ basis, this interaction sequence will
realize the C-NOT operations from cavity $1$ to cavity $3$:

\begin{widetext}
\begin{subequations}
\begin{equation}
|\alpha_{+}\rangle _{1}|e_{a}\rangle |\pm \alpha_{3}\rangle
\overset{\lambda t_{1}=\pi/2}{\longrightarrow }|\alpha _{+}\rangle
_{1}|e_{a}\rangle |\pm \alpha _{3}\rangle
\overset{R_{a}}{\longrightarrow } |\alpha _{+}\rangle
_{1}|+_{a}\rangle |\pm \alpha _{3}\rangle \overset{\lambda
t_{3}=\pi/2}{\longrightarrow }|\alpha _{+}\rangle
_{1}|+_{a}\rangle |\mp \alpha _{3}\rangle  \label{cnot3a}
\end{equation}
\begin{equation}
|\alpha_{-}\rangle _{1}|e_{a}\rangle |\pm \alpha_{3}\rangle
\overset{\lambda t_{1}=\pi/2}{\longrightarrow }|\alpha
_{-}\rangle_{1}|g_{a}\rangle |\pm
\alpha_{3}\rangle\overset{R_{a}}{\longrightarrow } |\alpha
_{+}\rangle _{1}|-_{a}\rangle |\pm \alpha _{3}\rangle \overset{
\lambda t_{3}=\pi/2}{\longrightarrow }|\alpha _{+}\rangle
_{1}|-_{a}\rangle |\pm \alpha _{3}\rangle  \label{cnot3b}
\end{equation}
\end{subequations}
\end{widetext}

The two mixed states to be purified are:
\begin{subequations}
\begin{equation}
\rho_{12}=F|\Phi^{+}\rangle_{12}\langle
\Phi^{+}|+(1-F)|\Psi^{+}\rangle_{12}\langle \Psi^{+}|
\label{rho12}
\end{equation}
\begin{equation}
\rho_{34}=F|\Phi^{+}\rangle_{34}\langle
\Phi^{+}|+(1-F)|\Psi^{+}\rangle_{34}\langle \Psi^{+}|
\label{rho34}
\end{equation}
\end{subequations}
where Eq (\ref{rho12}) is expressed in basis $\{|\alpha
_{+}\rangle_{1}|\alpha _{+}\rangle _{2}$, $|\alpha _{+}\rangle
_{1}|\alpha _{-}\rangle _{2}$, $|\alpha _{-}\rangle _{1}|\alpha
_{+}\rangle _{2}$, $|\alpha _{-}\rangle _{1}|\alpha _{-}\rangle
_{2}\}$, and Eq (\ref{rho34}) is expressed in basis $\{|\alpha
_{3}\rangle |\alpha _{4}\rangle$, $|\alpha _{3}\rangle |-\alpha
_{4}\rangle$, $|-\alpha _{3}\rangle |\alpha _{4}\rangle$,
$|-\alpha_{3}\rangle |-\alpha _{4}\rangle\}$. Here the fidelity of
the mixed state relative to the initial maximally entangled state
is $F=\langle \Phi ^{+}|\rho |\Phi ^{+}\rangle$.

After the procedure described in Eqs (\ref{cnot3a}, \ref{cnot3b}),
we can measure the atoms $a$, $b$. There will be four possible
results, $|e_{a}\rangle |e_{b}\rangle$, $|g_{a}\rangle
|g_{b}\rangle$, $|e_{a}\rangle |g_{b}\rangle$, and $|g_{a}\rangle
|e_{b}\rangle$. These four results can be divided into two kinds,
$|e_{a}\rangle |e_{b}\rangle$, $|g_{a}\rangle |g_{b}\rangle$ and
$|e_{a}\rangle |g_{b}\rangle$, $|g_{a}\rangle |e_{b}\rangle$.
Corresponding to each kind, we all can get the purified entangled
coherent state of cavity $1$, $2$ conditioned on the measurement
result on cavity $3$, $4$ in the $\{|\alpha _{3}\rangle |\alpha
_{4}\rangle$, $|\alpha _{3}\rangle |-\alpha _{4}\rangle$,
$|-\alpha _{3}\rangle |\alpha _{4}\rangle$, $|-\alpha _{3}\rangle
|-\alpha _{4}\rangle\}$ basis.

For the first kind of result $|e_{a}\rangle |e_{b}\rangle$,
$|g_{a}\rangle |g_{b}\rangle$, we can get the new state of cavity
fields $1$, $2$ provided that cavity fields $3$, $4$ are all in
the same coherent state.
\begin{subequations}
\begin{equation}
\rho _{12new}=F_{new}|\Phi^{+}\rangle _{12}\langle \Phi
^{+}|+(1-F_{new})|\Psi ^{+}\rangle _{12}\langle \Psi^{+}|
\label{rho12a}
\end{equation}
\begin{equation}
F_{new}=\frac{F^{2}}{F^{2}+(1-F)^{2}}.
 \label{fidelity1}
\end{equation}
\end{subequations}
For the second kind of result $|e_{a}\rangle |g_{b}\rangle
,|g_{a}\rangle |e_{b}\rangle $\ we can get that:
\begin{subequations}
\begin{equation}
\rho^{'}_{12new}=F^{'}_{new}|\Phi^{-}\rangle _{12}\langle \Phi
^{-}|+(1-F^{'}_{new})|\Psi ^{-}\rangle _{12}\langle \Psi^{-}|
\label{rho12b}
\end{equation}
\begin{equation}
F^{'}_{new}=F_{new}.
 \label{fidelity2}
\end{equation}
\end{subequations}

From the result in Eqs (\ref{rho12a}, \ref{rho12b},
\ref{fidelity1}, \ref{fidelity2}), we find that the mixed
entangled state in Eq (\ref{rho12}) has been purified through the
C-NOT operations form cavities $1$, $2$ to cavities $3$, $4$ plus
the measurements on atoms and cavities. Consider the first result
as example. When $F>\frac{1}{2}$, $F_{new}>F$. So the mixed state
in Eq (\ref{rho12}) has been purified. Because the initial
Fidelity $F$ is an arbitrary number between $0.5$ and $1.0$, the
iteration of the above scheme can extract a entangled coherent
state with an degree of entanglement arbitrarily close to $1.0$.
The same analysis applies to the second result.

In the above scheme, the C-NOT operations from one cavity to
another has been realized using the highly detuned interaction
between three-level $\Lambda$-type atoms and coherent optical
fields. In fact, using this kind of interaction we also can
realize the C-NOT operations from one atom to anther.

Let the first atom ($1$) through a cavity, initially prepared in
coherent state. The interaction is governed by the Hamiltonian
expressed in Eq (\ref{hamiltonian}). After the first atom flying
out of the cavity, we can complete the rotational operation on the
coherent state of the cavity:
$|\alpha\rangle\rightarrow|\alpha_{+}\rangle,
|-\alpha\rangle\rightarrow|\alpha_{-}\rangle$ by using nonlinear
Kerr medium~\cite{Yurke:1986}. Then we will sent the second atom
($2$) through the cavity. If the interaction times between the
cavity and atoms $1$, $2$ are all equal to $t=\pi/(2\lambda)$, the
total evolution of the system can be expressed as:

\begin{widetext}
\begin{subequations}
\begin{equation}
|+\rangle_{1}|\alpha\rangle|e_{2}\rangle\overset{\lambda
t_{1}=\pi/2}{\longrightarrow}|+\rangle_{1}|-\alpha\rangle|e_{2}\rangle\overset{R}{\longrightarrow}|+\rangle_{1}|\alpha_{-}\rangle|e_{2}\rangle\overset{\lambda
t_{2}=\pi/2}{\longrightarrow}|+\rangle_{1}|\alpha_{-}\rangle|g_{2}\rangle
\label{cnot4a}
\end{equation}
\begin{equation}
|+\rangle_{1}|\alpha\rangle|g_{2}\rangle\overset{\lambda
t_{1}=\pi/2}{\longrightarrow}|+\rangle_{1}|-\alpha\rangle|g_{2}\rangle\overset{R}{\longrightarrow}|+\rangle_{1}|\alpha_{-}\rangle|g_{2}\rangle\overset{\lambda
t_{2}=\pi/2}{\longrightarrow}|+\rangle_{1}|\alpha_{-}\rangle|e_{2}\rangle
\label{cnot4b}
\end{equation}
\begin{equation}
|-\rangle_{1}|\alpha\rangle|e_{2}\rangle\overset{\lambda
t_{1}=\pi/2}{\longrightarrow}|-\rangle_{1}|\alpha\rangle|e_{2}\rangle\overset{R}{\longrightarrow}|-\rangle_{1}|\alpha_{+}\rangle|e_{2}\rangle\overset{\lambda
t_{2}=\pi/2}{\longrightarrow}|-\rangle_{1}|\alpha_{+}\rangle|e_{2}\rangle
\label{cnot4c}
\end{equation}
\begin{equation}
|-\rangle_{1}|\alpha\rangle|g_{2}\rangle\overset{\lambda
t_{1}=\pi/2}{\longrightarrow}|-\rangle_{1}|\alpha\rangle|g_{2}\rangle\overset{R}{\longrightarrow}|-\rangle_{1}|\alpha_{+}\rangle|g_{2}\rangle\overset{\lambda
t_{2}=\pi/2}{\longrightarrow}|-\rangle_{1}|\alpha_{+}\rangle|g_{2}\rangle
\label{cnot4d}
\end{equation}
\end{subequations}
\end{widetext}
That is to say, we also realize the atom-to-atom C-NOT operations.

Then, because we have realized the C-NOT operations: atom-to-atom,
field-to-field, atom-to-field and field-to-atom, the teleportation
schemes for unknown coherent superposition state of cavity and
unknown atomic states can be easily realized~\cite{Bennett:1993}.
That is to say, the joint Bell state measurement will be carried
out by sending an atom through a detuned optical field, i.e. the
Bell state measurement will be converted into the product single
atom measurement and single field measurement.

The detection of coherent field has been realized by B.Yurke
\emph{et al}~\cite{Yurke:1986}, and they can distinguish
$|\alpha\rangle$ and $|-\alpha\rangle$ by homodyne detection. So
the detection of coherent field in our scheme is realizable, and
the detection of atom can be realized by the field-induced
ionization~\cite{Raimond:2001}.

We now consider the implementation of the above-mentioned scheme.
The atoms used in our scheme are all $\lambda$-type three level
atoms with one excited level and two degenerate ground levels,
which can be achieved by using Zeeman sublevels. From the
discussion in~\cite{zheng:song}, by using Rydberg atom of long
lifetime and superconducting microwave cavities with an enough
high-Q, there is sufficient time to achieve our schemes in
experiment.

In conclusion, Using the highly detuned interaction between
three-level $\lambda$-type atoms and coherent optical fields, the
C-NOT operations from atoms to atoms, optical fields to optical
fields, from atoms to optical fields and from optical fields to
atoms all have been realized in cavity QED. Our scheme not only
can generate multi-atom entangled states but also can generate
multi-mode entangled coherent states. Based on the C-NOT gates,
the entanglement purification for mixed entangled coherent states
and teleportation of unknown states of atoms or fields all have
been proposed. In the previous quantum information processing
proposals, many atoms are required to interact with single mode
field simultaneously. In our scheme, only the interaction between
single atom and single mode field is needed, which avoids the
problem of the synchronization of many atoms in the previous
quantum information processing proposals.

\begin{acknowledgements}
This work is supported by the Natural Science Foundation of the
Education Department of Anhui Province under Grant No: 2004kj005zd
and Anhui Provincial Natural Science Foundation under Grant No:
03042401 and the Talent Foundation of Anhui University.
\end{acknowledgements}

\end{document}